\documentclass{emulateapj}
\usepackage[below]{placeins}
\usepackage{natbib}
\slugcomment{accepted for publication by the Astrophysical Journal Letters}

\def\gtorder{\mathrel{\raise.3ex\hbox{$>$}\mkern-14mu
    \lower0.6ex\hbox{$\sim$}}}
\def\ltorder{\mathrel{\raise.3ex\hbox{$<$}\mkern-14mu
    \lower0.6ex\hbox{$\sim$}}}

\def\kmsmpc{\ {\rm km~s^{-1} Mpc^{-1}}}
\def\msun{M_\odot}

\newcommand {\rs} {$R_{\rm s}$}

\newcommand {\mvir} {$M_{\rm vir}$}

\newcommand {\rhos} {$\rho_{\rm s}$}

\shorttitle{Erasing Dark Matter Cusps with Baryons}
\shortauthors{Romano-Diaz et al.}

\begin{document}

\title{Erasing Dark Matter Cusps in Cosmological Galactic Halos 
with Baryons}

\author{ 
Emilio Romano-D\'{\i}az\altaffilmark{1},
Isaac Shlosman\altaffilmark{1},
Yehuda Hoffman\altaffilmark{2},
Clayton Heller\altaffilmark{3}
}
\altaffiltext{1}{
Department of Physics and Astronomy, 
University of Kentucky, 
Lexington, KY 40506-0055, 
USA
}
\altaffiltext{2}{
Racah Institute of Physics, Hebrew University; Jerusalem 91904, Israel
}
\altaffiltext{3}{
Department of Physics, 
Georgia Southern University, 
Statesboro, GA 30460, 
USA
}

\begin{abstract}
We study the central dark matter (DM) cusp evolution in cosmologically grown 
galactic halos. Numerical models with and without baryons (baryons$+$DM, hereafter 
BDM model, and pure DM, PDM model, respectively) are advanced from identical 
initial conditions, obtained using the Constrained Realization method.
The DM cusp properties are 
contrasted by a direct comparison of pure DM and baryonic models. We find a 
divergent evolution between the PDM  and BDM models within the inner 
few$\times 10$~kpc region. The PDM model forms a $R^{-1}$ cusp as expected,
while the DM in the BDM model forms a larger isothermal cusp $R^{-2}$ instead. The
isothermal cusp is stable until $z\sim 1$ when it gradually levels off. This
leveling proceeds from inside out and the final density slope is shallower
than $-1$ within the central 3~kpc (i.e., expected size of the $R^{-1}$ cusp), 
tending to a flat core within $\sim 2$~kpc. This effect cannot be explained
by a finite resolution of our code which produces only a 5\% difference
between the gravitationally softened force and the exact Newtonian force of point 
masses at 1~kpc from the center. Neither is it related to the energy feedback 
from stellar evolution or angular momentum transfer from the bar. Instead
it can be associated with the action of DM$+$baryon subhalos heating up the cusp
region via dynamical friction and forcing the DM in the cusp to flow out and 
to `cool' down. The process described here is not limited to low $z$ and can
be efficient at intermediate and even high $z$.
\end{abstract}

    
\section{Introduction}
\label{sec:intro}

Cosmological numerical simulations of pure dark matter (DM) halo formation 
have led to ``universal" density profile which can be approximated
by $\rho(r) = 4 \rho_{\rm s}/(r/R_{\rm s})(1+r/R_{\rm s})^2$ (Navarro et al. 1997, NFW),
with \rs\ being a characteristic ``inner" radius where
logarithmic density slope is $-2$ and \rhos\ is the density at \rs. This
profile steepens outside \rs\ and tends to a cusp $\sim R^{-1}$ toward the center,
at $R_{\rm cusp}\sim 0.1 R_{\rm s}$.
While this profile remains invariant under mergers (El-Zant 2008), its origin is yet
to be explained (e.g., Syer \& White 1998; Nusser \& Sheth 1999; Shapiro et al. 2004; 
Ascasibar et al. 2007)

The NFW density profile is a source of an ongoing controversy --- its universality
in the numerical pure DM models is in apparent contradiction with at least some
of the observations of disk galaxies and galaxy clusters, which exhibit rather
flat density cores (e.g., Flores \& Primack 1994; Kravtsov et al.
1998; Salucci \& Burkert 2000; Sand et al. 2002; de Blok et al. 2003; de Blok 2005, 
2007; Simon et al. 2003; but see Rhee et al. 2004; answered by Gentile et al. 2004, 2005;
de Naray et al. 2008).  
While a number of `exotic' explanations involving less conventional physics
have been proposed, attempts to resolve this discrepancy within
the CDM cosmology have been made as well. The DM cusp leveling off was
attributed to the DM-baryon interactions, such as a dynamical friction of
DM/baryon inhomogeneities (i.e., substructure) against the DM halo background 
(El-Zant et al. 2001, 2004; Tonini et al. 2006), stellar bar--DM interaction (Weinberg 
\& Katz 2002; Holley-Bockelmann et al. 2005; but see Sellwood 2003; McMillan \& Dehnen 
2005; Dubinski et al. 2008), and baryon energy feedback (e.g., Mashchenko et al. 2006; 
Peirani et al. 2008). 
In this Letter, we test the first option (El-Zant et al.) --- the effect of
the baryon$+$DM substructure in the fully self-consistent numerical simulations of 
halo formation in the $\Lambda$CDM cosmology with WMAP3 paramaters (more details in 
Romano-D\'{\i}az et al. 2008). 

Early arguments about change of the central density profile claimed
that baryons drag DM in the so-called adiabatic contraction, steepening the 
DM density slope (e.g., Blumenthal et al. 1986). However, they have  
neglected the clumpy nature of the accreting material in the assembling halo,
which can give rise to clump-background dynamical friction and
energy transfer from the clumps to the background. The ability of baryons to radiate 
their internal energy increases the binding energy and acts as a `glue' on the DM 
substructure (El-Zant et al. 2001). El-Zant et al. have used the Monte-Carlo technique 
to calculate the leveling of the NFW cusp in the presence of such clumps and found this 
process to be efficient over $\sim 1-2$~Gyr. The initial conditions
presumed the existence of a NFW cusp, a spherically-symmetric DM halo and 
indestructible clumps.

\begin{figure*}[ht!!!!!!!!!!!!!!!!!!!]
\begin{center}
\includegraphics[angle=0,scale=0.28]{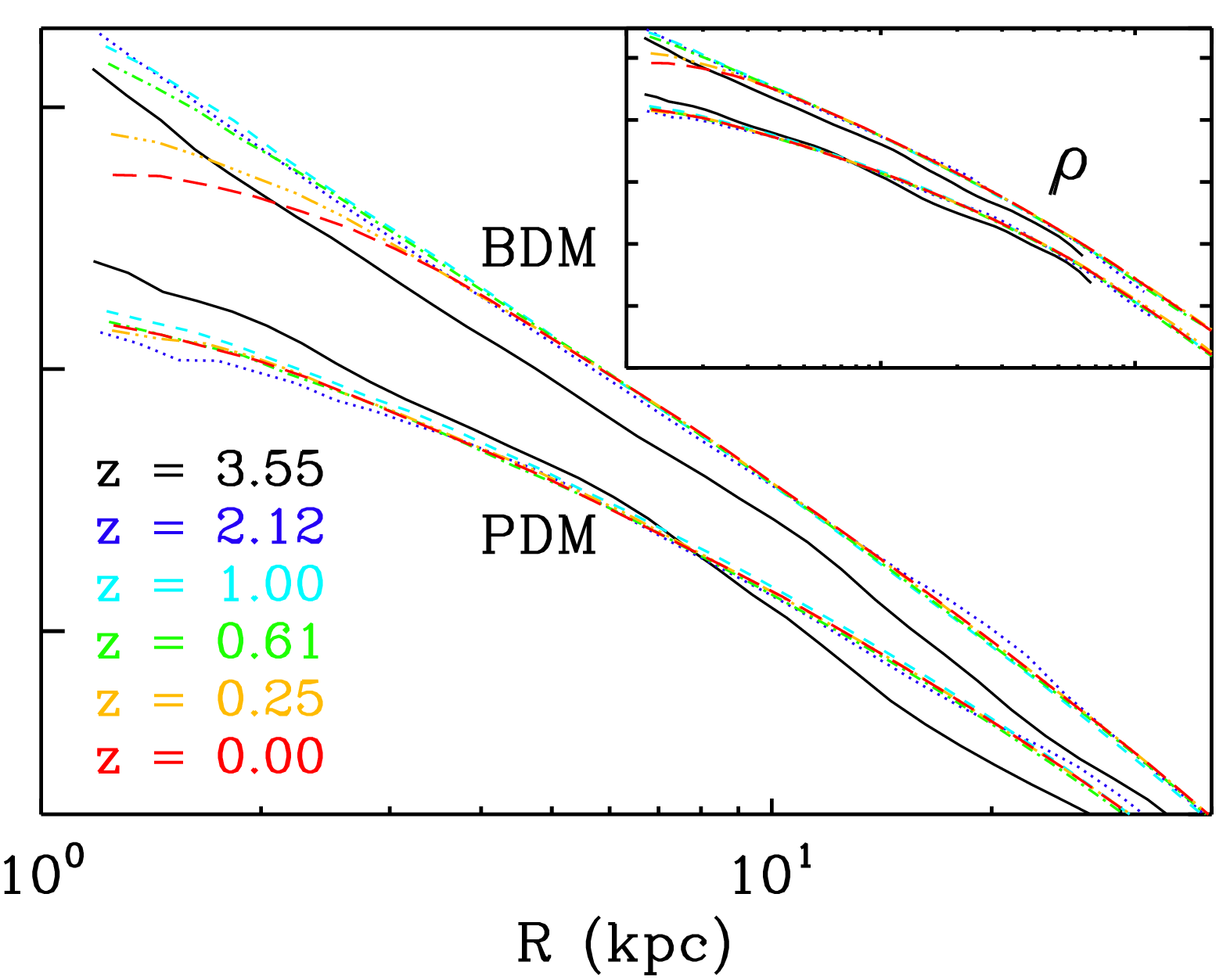}
\includegraphics[angle=0,scale=0.28]{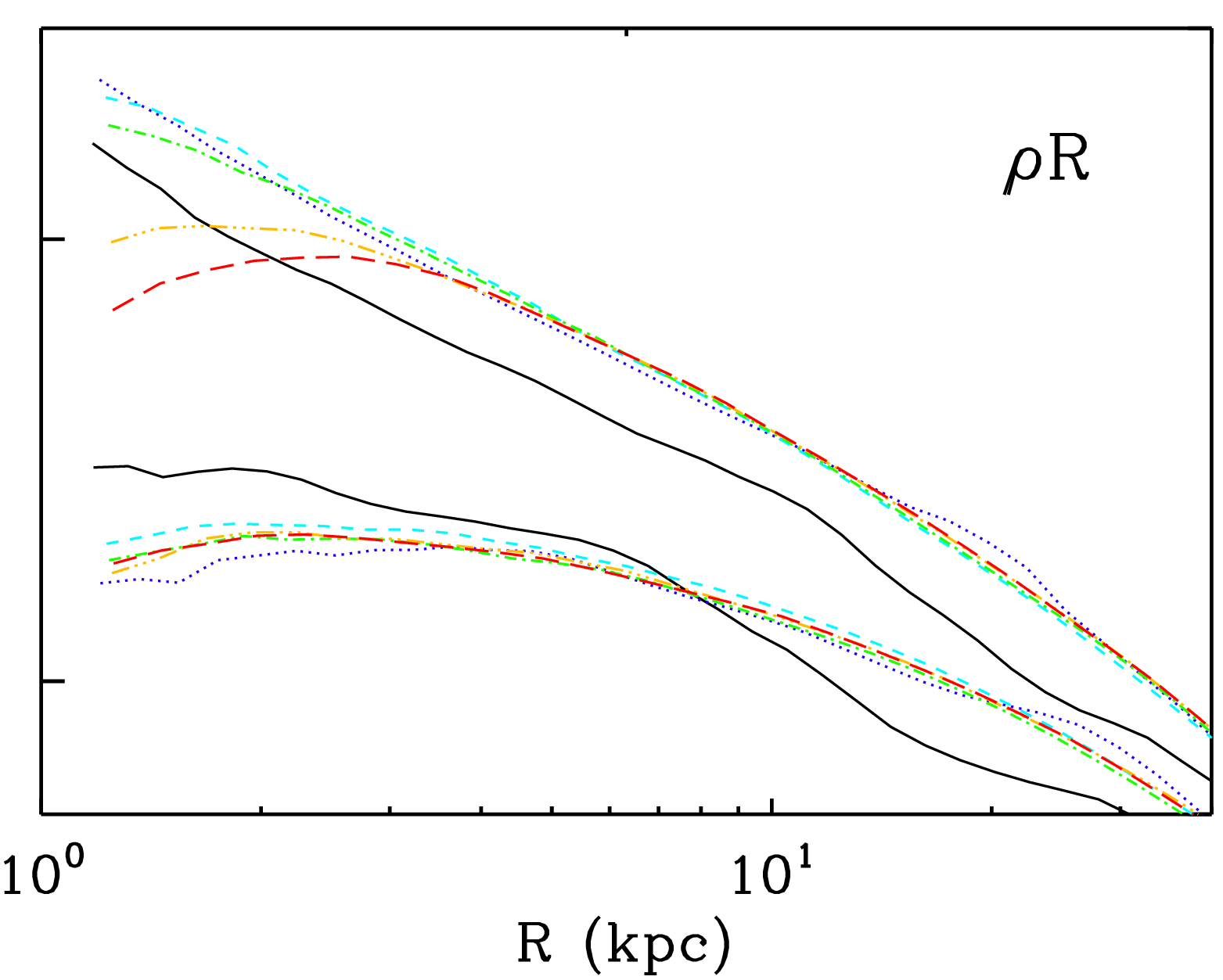}
\includegraphics[angle=0,scale=0.28]{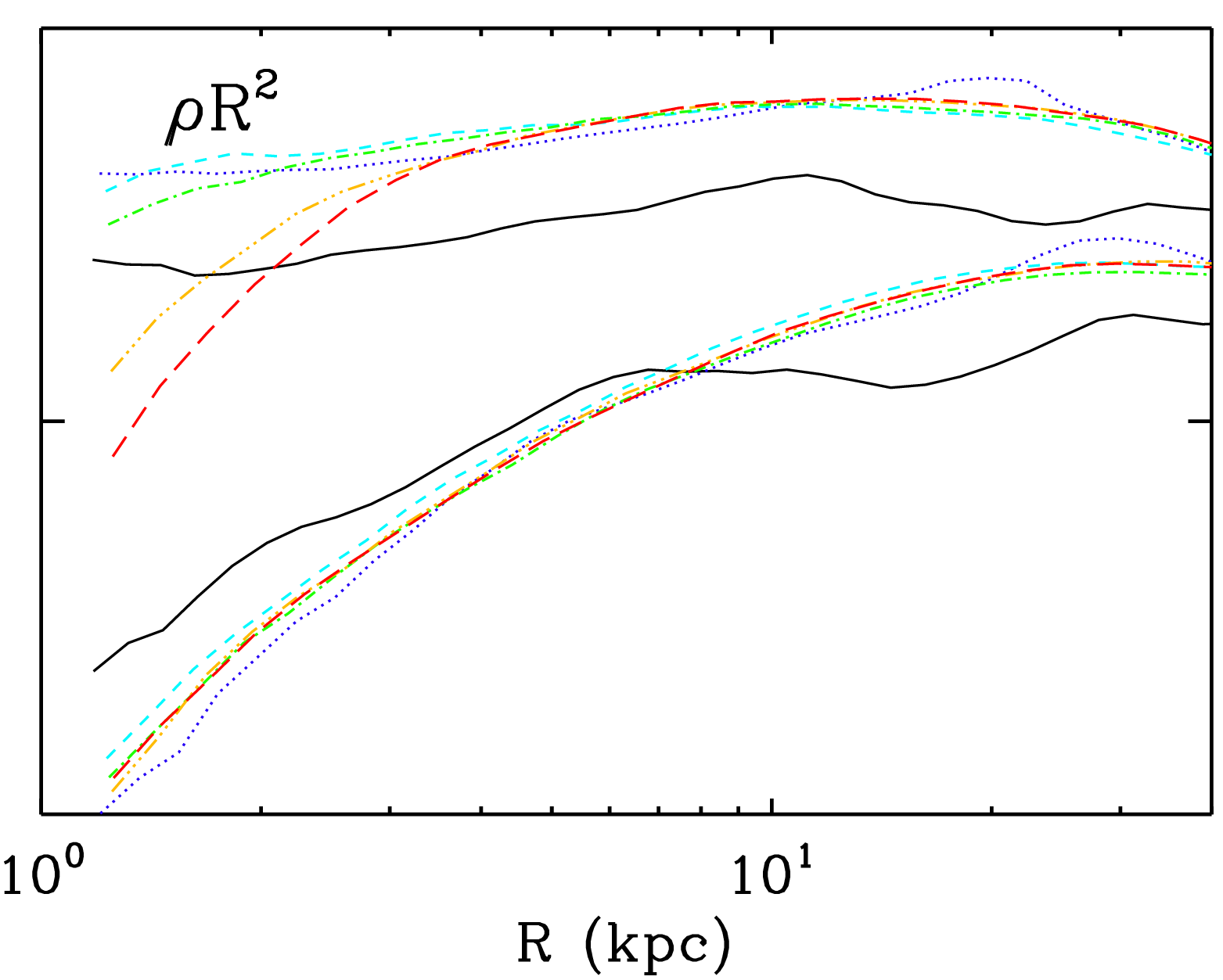}
\end{center}
\caption{Redshift evolution of DM density profiles, $\rho(R)$ (left), 
$\rho(R) R$ (middle) and $\rho(R) R^2$ (right) in PDM and BDM models: $z=3.55$ (solid),
2.12 (dotted), 1.0 (dashed), 0.61 (dot-dashed), 025 (dot-dash-dotted) and 0 (long
dashed). The PDM
and BDM curves are displaced vertically for clarity. The inner 40~kpc of halos are 
shown. The vertical coordinate units are logarithmic and arbitrary. For the PDM model, 
the density is well fitted by the NFW profile over a large range in $z$, and 
$R_{\rm s}\sim 28$~kpc at $z=0$. For the BDM model, the NFW fit is worse and
$R_{\rm iso}\sim 15$~kpc at the end. The insert provides $\rho$ within 200~kpc
range for a comparison.
}
\end{figure*}

The NFW cusp is characterized by the 
``temperature" (i.e., DM dispersion velocities) inversion which makes it 
thermodynamically improbable but dynamically stable in the absence of energy 
transfer mechanism (El-Zant et al. 2001). Dynamical friction of sufficiently 
bound clumps can trigger such energy flux into the cusp, increasing its dispersion 
velocities and washing it out. El-Zant et al. found that the necessary 
condition for the dynamical friction 
to have an effect is the aggregation of mass into the clumps more massive than 
$\sim 10^{-4}$ of the DM halo --- result confirmed for the galaxy clusters as well 
(El-Zant et al. 2004). 

In fully self-consistent numerical simulations with baryons, the question
is whether the NFW cusp forms in the first place. Such simulations have been 
either limited to high redshifts ($z=3.3$, Gnedin et al. 2004) or focused on the 
adiabatic contraction and had an insufficient resolution (Gustafsson et al. 2006).  
They have verified that baryons lead to a steeper DM density profile than the NFW
cusp, but left open its subsequent evolution. This issue is addressed here.

\section{Numerics and Initial Conditions}

Numerical simulations have been performed with the FTM~4.5 hybrid $N$-body/SPH 
code (Heller \& Shlosman 1994; Heller et al. 2007; 
Romano-D\'{\i}az et al. 2008) using physical and not comoving coordinates. The 
number of DM particles is $2.2\times 10^6$
and the SPH particles --- $4\times 10^5$. The gravity is computed with the falcON 
routine (Dehnen 2002) which scales as O(N). The gravitational softening (P1 
of falcON) is $\epsilon_{\rm grav}= 500$~pc, for DM, stars and gas. The calculated 
force differs by $\sim 5\%$ from the exact 
Newtonian force between point masses at $R\sim 2\epsilon_{\rm grav} = 1$~kpc.
We assume the $\Lambda$CDM 
cosmology with WMAP3 parameters, $\Omega_{\rm m}=0.24$, $\Omega_\Lambda=0.76$
and $h=0.73$, where $h$ is the Hubble constant in units of $100~\kmsmpc$. The
variance $\sigma_8=0.76$ of the density field convolved with the top hat window
of radius $8h^{-1}$~Mpc$^{-1}$ was used to normalize the power spectrum.
The star formation algorithm is described in Heller et al. (2007). 
Several generations of stars are allowed to form from an SPH particle. 
The energy and momentum feedback into the ISM is
implemented, and the relevant parameters are (Heller et al.): the energy 
thermalization $\epsilon_{\rm SF}=0.3$, the cloud gravitational collapse 
$\alpha_{\rm ff}=1$, and the self-gravity fudge factor $\alpha_{\rm crit}=0.5$. 

The initial conditions generated here are those of Romano-D\'{\i}az et al. (2008): 
we use
the Constrained Realizations (CR) method (Hoffman \& Ribak 1991) within a restricted
volume of $8^3h^{-1}$~Mpc, where a $5h^{-1}$~Mpc sphere is carved out and evolved. 
The constructed Gaussian field is required to obey a set of constraints
of arbitrary amplitudes and positions (e.g., Romano-D\'{\i}az
et al. 2006, 2007).  
Two constraints were imposed  on the initial density field, first ---
that the linear field Gaussian smoothed with a
kernel  of $1.0\times 10^{12}~h^{-1}\msun$ has an over-density of $\delta=3$ at 
the origin ($2.5\sigma$ perturbation, where $\sigma^2$ is the variance of the 
appropriately smoothed field). It was imposed on a $256^3$ grid 
and predicted to collapse by $z_{\rm c}\sim 1.33$ 
by the top-hat model. This perturbation is embedded in a region (2nd constraint) 
corresponding to  
$5\times 10^{13}~h^{-1}\msun$ in which the over-density is zero, i.e.,
the unperturbed universe. The random component of the CR introduces density
perturbations on all scales, thus leading to major mergers (Romano-D\'{\i}az et al. 
2006, 2007).
The mass inside the computational sphere is $\sim 6.1\times 10^{12}~h^{-1}\msun$. 
In the baryonic model, we have randomly replaced 1/6 of DM particles by equal mass 
SPH particles. Hence, $\Omega_{\rm m}$ is not affected.

\section{DM Density and Velocity Dispersion Profiles}

Two models analyzed here are those of a pure DM (hereafter PDM) and DM$+$baryons
(BDM). Evolution of DM density profiles for both models is shown in Fig.~1
at various redshifts and in three different ways: as $\rho(R)$ {\it per se}, factored 
by $R$, and by $R^2$, in order to emphasize the $R^{-1}$ and $R^{-2}$
cusps. At early redshifts, $z\sim 7$, the DM density profiles of both models are 
nearly identical. With time, the BDM halo becomes more cuspy --- the DM experiencing 
the adiabatic contraction by baryons. This region of higher density extends gradually to 
larger $R$, up to $\sim 15$~kpc at $z=0$. Between $z\sim 4$ and 1 the 
DM density profile becomes and remains nearly isothermal within this region --- 
$\rho(R)R^2$ is flat 
there (Fig.~1, right frame). In comparison, the PDM halo exhibits the slope of $-2$ 
at $R_{\rm s}\sim 28$~kpc only, as expected, with the NFW cusp size 
$R_{\rm cusp}\sim 3$~kpc at $z=0$. Hence over prolonged time period, while the PDM 
model displays the NFW cusp, the DM in the BDM model remarkably forms an isothermal 
cusp, down to the central kpc, i.e., to $\sim 2\epsilon_{\rm grav}$. 
The isothermal cusp in the BDM model is gradually erased inside $R_{\rm iso}\sim 15$~kpc
after $z\sim 1$. In fact, within the central 3~kpc it becomes flatter than $-1$,
i.e., flatter than the NFW cusp, as seen in all three frames of Fig.~1, and within 
$\sim 2$~kpc forms a rather flat core. At $z=0$, the PDM model is denser outside
$R_{\rm iso}$, while BDM remains denser within this radius. 

\begin{figure} 
\begin{center}
\includegraphics[angle=0,scale=0.47]{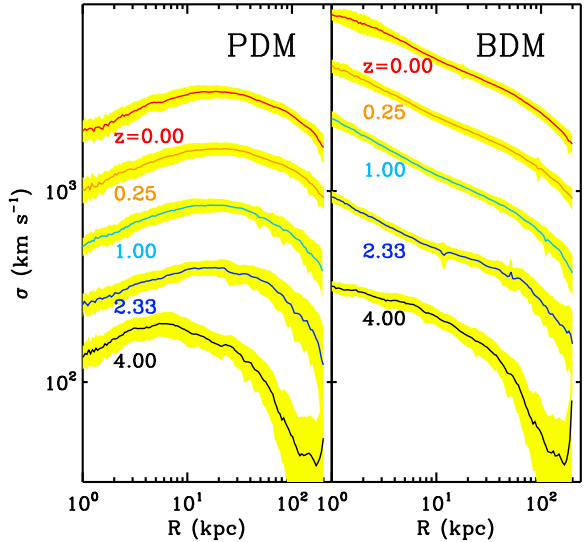}
\end{center}
\caption{Redshift evolution of DM velocity dispersions in PDM (left) and BDM (right) 
models. Except for the lowest ones, the curves are displaced vertically up for clarity. The
second curves (from the bottom) are displaced by a factor of 2, the third 
--- by a factor of $2^2$, the fourth --- by a factor of $2^3$, and the last ones --- by
a factor of $2^4$. The colored width represents a 1$\sigma$
dispersion around the mean. The inner 200~kpc of halos are shown. 
The vertical coordinate units are logarithmic. 
}
\end{figure}

Furthermore, in the PDM model, the density around 1~kpc from the center stays 
virtually unchanged
after $z\sim 3$. The baryonic model shows a somewhat different behavior. Namely,
it keeps the same DM density at 1~kpc between 
$z\sim 3-0.7$, which decays thereafter by a factor of $\sim 3$. 

Behavior of the velocity dispersions, $\sigma_{\rm DM}$, in both models mirrors 
that of the density.
The NFW DM cusp in the PDM model forms early and is characterized by the `temperature'
inversion as shown in Fig.~2, where we display the $z$ evolution of the DM dispersion 
velocities (see also El-Zant et al. 2001). In contrast, the BDM model shows a cuspy
distribution of $\sigma_{\rm DM}(R)$, which after $z\sim 4$ can be approximated well by
a single power law, $\sigma_{\rm DM}(R)^2\sim R^{-\beta}$ inside central 10~kpc. The 
evolution of $|\beta|$ is shown
in Fig.~3 for both models. It is steadily increasing in the PDM model until
the end of the major mergers epoch at $z\sim 1.5$, where it levels off. The BDM model
shows a similar increase in $|\beta|$, but well beyond the mergers epoch, until 
$z\sim 0.7$, where it sharply declines. This decline in $\beta$ is crucial in 
understanding the fate of the density cusp, as we discuss below.

\begin{figure}
\begin{center}
\includegraphics[angle=0,scale=0.56]{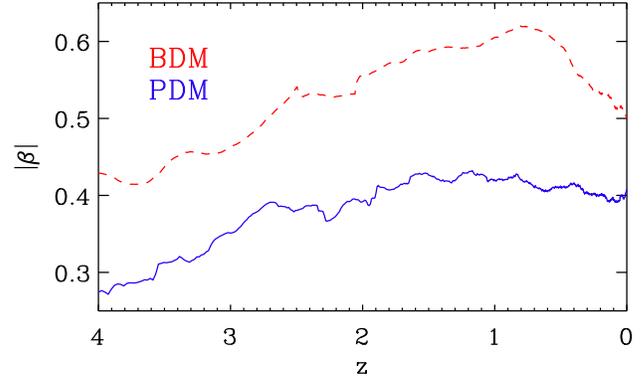}
\end{center}
\caption{Redshift evolution of the power index $|\beta|$ in the PDM (blue solid) and 
BDM (dashed red) models. $\beta$ is calculated by fitting a power law 
$\sigma_{\rm DM}^2\sim R^{-\beta}$ between 1~kpc and 10~kpc. Note that $\beta>0$ for
BDM and $\beta<0$ for PDM models.
}
\end{figure}

\section{Discussion}

We have performed a direct comparison between the DM halo evolution in the pure 
DM and baryonic models, from identical cosmological initial conditions. Our goal is 
to understand the effect of the baryons on the DM density profile. We, therefore,
focus on the DM evolution within the central region, quantify
the cuspiness of the density distribution there, and attempt to relate its
evolution to that of the DM substructure.
Our results show that the DM in the PDM model has established a NFW cusp early, and
the only subsequent changes we observe are related to sudden increases in \rs\ at
the time of the major mergers (e.g., Romano-D\'{\i}az et al. 2006, 2007). On the other hand,
the baryonic (BDM) model goes through a two-step process: first, in the central 
$\sim 10$~kpc, the DM is dragged in by the baryons in what fits the adiabatic 
contraction (e.g., Gnedin et al. 2004; Gustafsson et al. 2006). This stage is 
concurrent with the formation
and growth of a galactic disk, but we did not verify whether the disk is solely 
responsible for the DM contraction. For an extended period of time, after 
$z\sim 4$, an isothermal cusp forms with $\rho_{\rm DM}\sim R^{-2}$
and persists. In the next step, the DM cusp is being gradually washed out
from inside out, mainly after $z\sim 1$. By $z=0$, the isothermal cusp
is largely erased, and the density slope within the central 10~kpc is less
steep than $-2$. In fact within the central $\sim 3$~kpc, it is less steep than 
the NFW slope of $-1$. We note, that the {\it DM in the BDM model does not form the
NFW cusp in the first place, but rather by-passes it in favor of an isothermal DM
cusp which is subsequently washed out.} For example, the `temperature' inversion
characteristic of the NFW cusp is never observed in the BDM model (Fig.~2).

We first argue, that this leveling of the DM cusp is not a numerical artifact,
i.e., is not caused by a finite resolution of the code which is  
$2\epsilon_{\rm grav}=1$~kpc. At this distance, softened and exact Newtonian 
forces between point masses differ by 5\%  --- this cannot account for the observed 
changes in density profile
further out. We also note that the DM density peak in the BDM model is occasionally offset
from the baryonic peak by a few kpc, around $z\sim 1$ and especially 0.4, with the 
subhalos influx into the center, in what can be treated as $m=1$ instability.
Therefore, we have fitted the DM halo profile based on the position of 
the DM density peak at each frame. The cusp will be `smeared' if this procedure 
is not followed (McMillan \& Dehnen 2005). Also, the prime halo moves
with respect to the center of mass of the computational sphere, this is related
to the residual large-scale streaming motions in the DM (Romano-D\'{\i}az et al. 2008). 

\begin{figure}[ht!!!!!!!!!!!!!!!!!!!!!!!!!!!!!]
\begin{center}
\includegraphics[angle=0,scale=0.56]{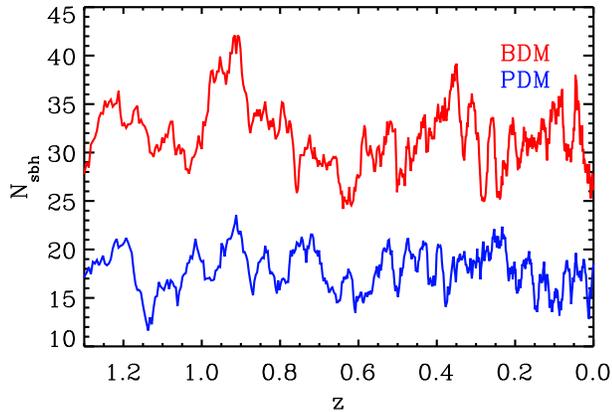}
\end{center}
\caption{Evolution of the subhalo number $N_{\rm sbh}$ within central 30~kpc
of the prime halo with redshift --- PDM (lower blue) and BDM (upper red). Note
the sharp increase in $N_{\rm sbh}$ for the BDM model after $z\sim 1$, 0.5
and 0.2. 
}
\end{figure}

Next, we note that most of the cusp leveling happens after $z\sim 1$ (Fig.~1). 
This correlates with a decrease in the slope of the central dispersion
velocities (Fig.~2). The slope change in $\beta$ (Fig.~3) comes 
from the decrease in the central dispersion velocities around few kpc, while
$\sigma_{\rm DM}$ at $R\sim 10$~kpc remains unchanged. What is the reason for 
this apparent `cooling' of the isothermal DM cusp?

Any decrease in central $\sigma_{\rm DM}$ of a dissipationless self-gravitationg 
entity in a
virial equilibrium is counter-intuitive. It clearly has nothing to do with the
major mergers, as this epoch ends at $z\sim 1.5$ in our models. Any subsequent
{\it smooth} accretion of dissipative baryons will only increase $\sigma_{\rm DM}$
as a result of the ongoing adiabatic contraction. However, if one considers
accretion of {\it clumpy} baryons, which act as a `glue' on the DM subhalos where
they reside, additional process of `heating' the DM by means of a dynamical friction
becomes important. The heating rate is skewed to the density peak and
will cause the DM to stream out of the
cusp's gravitational well, eventually decreasing the `temperature' gradient at the
center. Fig.~3 displays this behavior of $\beta$ after it reaches its peak at
$z\sim 0.7$. For a comparison, the PDM halo shows a flat $\beta$ after the
epoch of major mergers, for nearly 10~Gyr. 

Central region crossing by subhalos (as well as accretion and tidal disruption)
continues to the present time. Why do changes in the cusp become significant 
after $z\sim 1$? Fig.~4 displays the evolution of the subhalo population within 
the central 30~kpc of the BDM halo. 
We observe that at least in two instances, $z\sim 1$ and 0.4, there
is a clear excess in the BDM subhalo population, corresponding to the splash in the
subhalo influx rates. The number of these subhalos also exceeds that of
the PDM by $\sim 2$. The BDM subhalos appear in waves within
the central region and these waves coinside with the DM streaming out of the
center and `cooling' down. Hence, the process of washing out the DM cusp correlates 
with the influx of subhalos into the innermost region of the BDM halo. 
We show elsewhere that waves of subhalos crossing the central region of the
prime halo originate in the filament, cluster there and enter the prime halo
before they merge among themselves. 

We find that the mass distribution of subhalos (in the computational sphere) 
with masses $M_{\rm sbh}$ 
evolves with redshift --- at high $z\gtorder 4$ it can be approximated by a 
power law, $N_{\rm sbh}\sim M_{\rm sbh}^{-1}$, both in PDM and BDM models
(Romano-Diaz et al. 2008). But in the central 
virialized 100~kpc of the halo it evolves differently from the field subhalos.
Specifically, it is heavily skewed toward more massive ones after 
$z\sim 1$, while that of the PDM subhalos is only weakly so. This is significant, 
because the efficiency of dynamical friction 
increases substantially with the clump-to-prime halo virial mass ratio (El-Zant et al. 
2001). At 
$z\sim 4$, this ratio for the most massive subhalo residing within the virialized
halo is $\sim 0.06$. By $z\sim 1$ this ratio is $\sim 10^{-3}$, and at $z=0$ it is
$\sim 8\times 10^{-4}$ --- still higher than the minimum required by El-Zant et al. 
 
The size of the obtained core in the DM distribution of the BDM model can be
compared with observationally inferred cores. Following Salucci et al. (2007), for
the virial mass of our halo at $z=0$, \mvir $\sim 4\times 10^{12}~\msun$, 
the expected core is $\sim 1.6$~kpc. Spano et al. (2008) lists a large range of
core sizes, starting from $\sim 1$~kpc. Both are compatible with the value
obtained here.  

In summary, we find that the DM density cusp corresponding to $R_{\rm cusp}$ in 
the PDM model is leveled off by the action of subhalos in the presence of baryons 
in the BDM model. The energy feedback from stellar evolution has 
decayed by this time without
any effect on the cusp. We do not detect the angular momentum ($J$) transfer
from the bar to the cusp --- 
the latter $J$ stays constant in time. We also find that the BDM model forms a 
steeper isothermal rather 
than NFW-type cusp for extended time period. It goes through a two step process which
involves gravitational contraction, followed by the influx of subhalos which
heat up and dissolve the cusp. 

Our models, although containing about $1.6\times 10^6$ particles per halo, are still 
insufficient to fully resolve the formation of substructure around the prime halo.  
They should be viewed rather as a lower limit on the efficiency of the process 
reported here. Neither is the process described here limited to low $z$ --- one can
easily envision scenario(s) when the subhalos act at intermediate and even high $z$. 
Overall, this work stands in agreement with Gnedin et al. (2004) who stopped the 
simulation at redshift $z=3.3$, well before the leveling of the density cusp. 
However, Gustafsson et al. (2006) barely lacked resolution to detect the
cusp behavior observed here. 
We have recently learned that J.P. Ostriker (priv. comm.) obtained similar results
on the cusp flattening.

\acknowledgements
We are grateful to our colleagues for illuminating discussions
and comments. I.S. acknowledges a partial support by NASA and STScI. 



\end{document}